\documentclass[final]{svjour3}
\usepackage[dvipdfmx]{graphicx}
\usepackage{rotating}
\usepackage{amssymb}
\usepackage{mathptmx}
\usepackage[numbers]{natbib}
\makeatletter
\journalname{Journal of Low Temperature Physics}
\bibpunct{}{}{,}{s}{}{,}

\usepackage[nolist]{acronym}  % acronyms
\usepackage[usenames, dvipsnames]{color}

\newcommand{\planck}{{\it Planck}}
\newcommand{\bicep}{BICEP2}
\newcommand{\bicepkeck}{BICEP2/Keck Array}
\newcommand{\pbsa}{POLARBEAR/Simons Array}
\newcommand{\pb}{POLARBEAR}

\newcommand{\sptg}{SPT-3G}
\newcommand{\spider}{SPIDER}
\newcommand{\ebex}{EBEX}
\newcommand{\bmode}{{\it B-}mode}

\begin{document}

\newcommand{\hdblarrow}{H\makebox[0.9ex][l]{$\downdownarrows$}-}
\title{Developments of highly-multiplexed, multi-chroic pixels for Balloon-Borne Platforms}

\author{F. Aubin$^a$, S. Hanany$^a$, B. R. Johnson$^b$, A. Lee$^c$, A. Suzuki$^{c, d}$, B. Westbrook$^c$, and K. Young$^a$.}

\institute{$^a$ School of Physics and Astronomy, University of Minnesota, Minneapolis, MN 55455, USA;\\% Tel.: (612) 624-0673\\% Fax:\\
$^b$ Department of Physics, Columbia University, New York, NY, 10027, USA;\\
$^c$ Department of Physics, University of California, Berkeley, Berkeley, CA 94720, USA;\\
$^d$ Lawrence Berkeley National Lab, Berkeley, CA 94720, USA;\\
\email{faubin@umn.edu}}

\maketitle

\begin{abstract}

We present our work to develop and characterize low thermal conductance bolometers that are part of sinuous antenna multi-chroic pixels (SAMP).
We use longer, thinner and meandered bolometer legs to achieve 9~pW/K thermal conductance bolometers.
We also discuss the development of inductor-capacitor chips operated at 4 K to extend the multiplexing factor of the frequency domain multiplexing to 105, an increase of 60\% compared to the factor currently demonstrated for this readout system.
This technology development is motivated by EBEX-IDS, a balloon-borne polarimeter designed to characterize the polarization of foregrounds and to detect the primordial gravity waves through their \bmode~signature on the polarization of the cosmic microwave background.
EBEX-IDS will operate 20,562 transition edge sensor bolometers spread over 7 frequency bands between 150 and 360~GHz. 
Balloon and satellite platforms enable observations at frequencies inaccessible from the ground and with higher instantaneous sensitivity.
This development improves the readiness of the SAMP and frequency domain readout technologies for future satellite applications. 

\end{abstract}

\section{Introduction}

\begin{figure}[htbp]
\begin{center}
\includegraphics[height=4.8 cm, keepaspectratio]{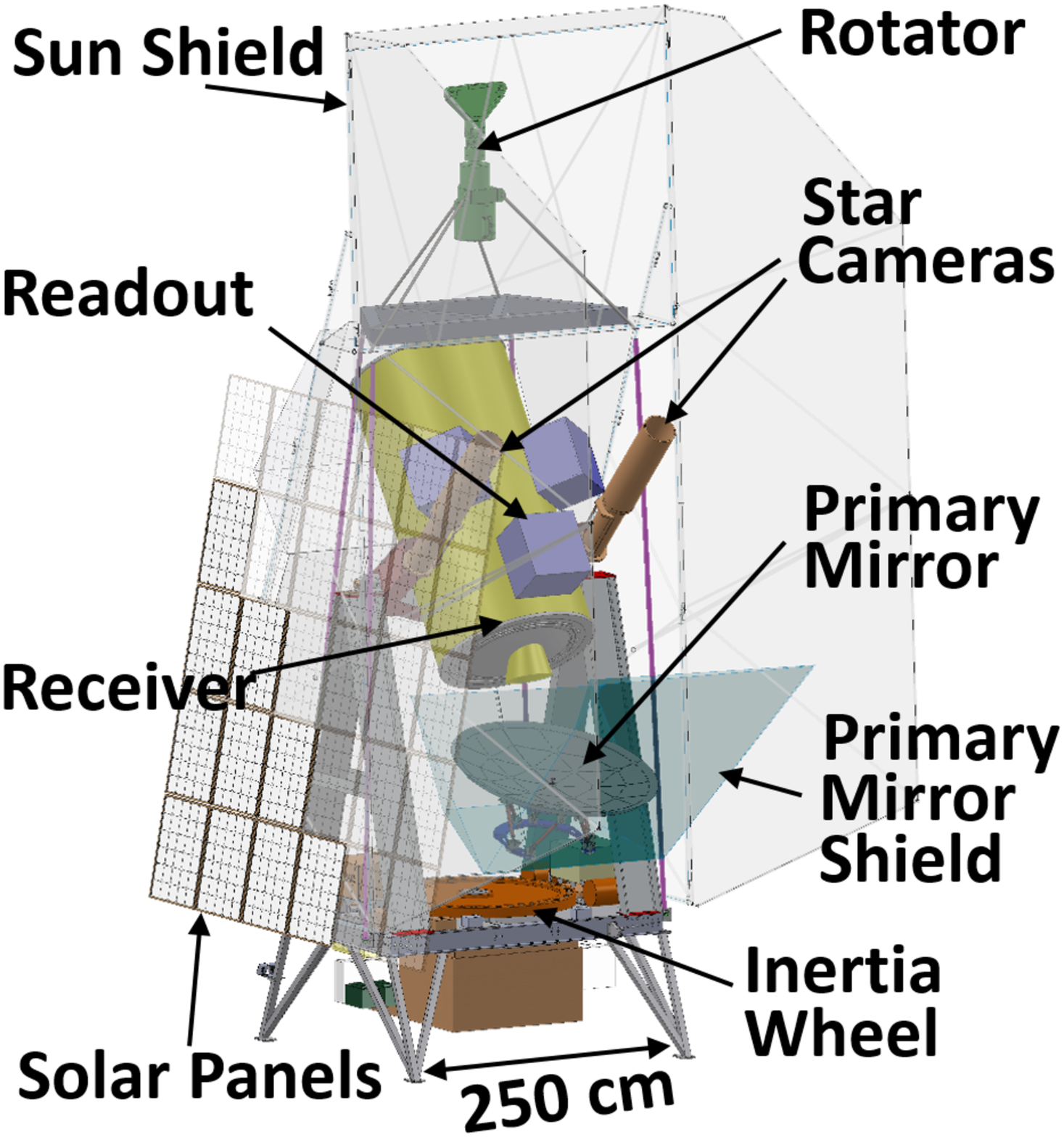}%\hspace{0.05\linewidth}
\includegraphics[height=4.8 cm, keepaspectratio]{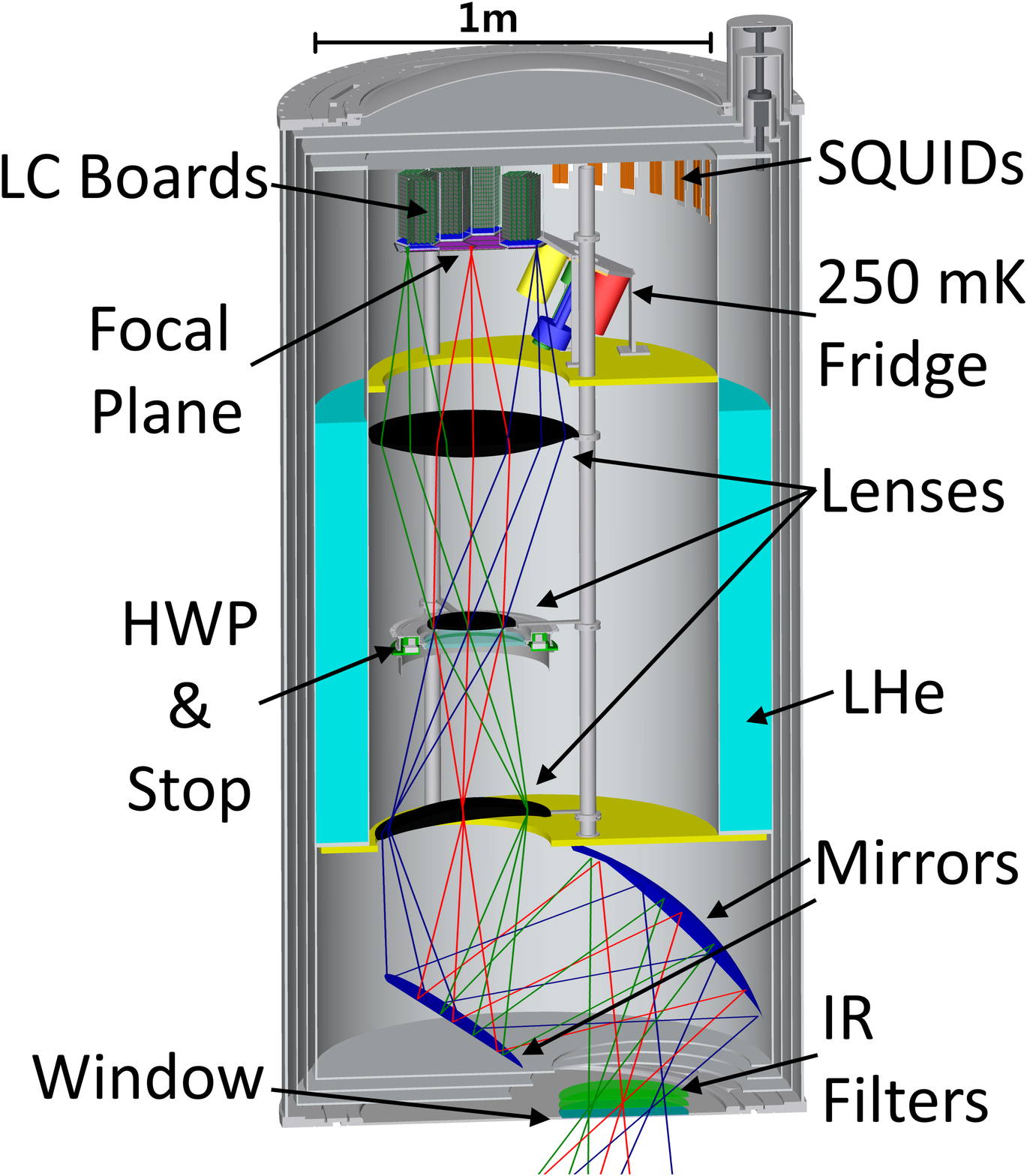}%\hspace{0.05\linewidth}
\caption{
{\it Left}: model of the \ac{EBEX-IDS} gondola.
{\it Right}: cut-through of a model of the \ac{EBEX-IDS} cryostat.
% {\it Right}: the 1500~deg$^2$ patch of sky \ac{EBEX-IDS} is planning to scan. 
}
\label{fig:ebexIdsDesign}
\end{center}
\end{figure}

The recent combined results from \planck~and \bicep~have shown that in order to detect the signature of primordial gravity waves as \bmode s~on the \ac{CMB}, the Galactic foregrounds must be thoroughly understood~[1].
The \ac{EBEX-IDS} is a balloon-borne polarimeter designed to achieve such detection and characterization.
\ac{EBEX-IDS} will re-use the attitude control system and most hardware components from \ebex~to mitigate operational risks.
Due to the lower density of atmosphere at float, the power absorbed by the detectors is at least a factor of 10 smaller compared to ground-based telescopes and is dominated by the instrument.
% ~2 k (T*opacity) dnu epsilon = 2(1.38e-23)(20 K)(40e9 Hz)(0.3) = 6.6 pW (240K * 0.5 = 20K) 
The 1.5 m aperture Gregorian Mizuguchi-Dragone telescope will be composed of an ambient temperature primary mirror and two additional mirrors cooled to 4~K to minimize instrument load on the detectors; see Figure~\ref{fig:ebexIdsDesign}.
The scanned patch 
will be shared with ground-based experiments such as \pbsa~and \bicepkeck~to improve the combined sensitivity while taking advantage of the deep coverage of these ground-based experiments and the high frequency coverage of the balloon-platform.
\ac{EBEX-IDS} will observe the sky 
% at high resolution 
in 7 different frequency bands from 150 to 360~GHz.

% \begin{table}[htbp]
%   \begin{center}
%     \begin{tabular}{|c|c|c|c|c|c|c|c|} \hline
%       Band (GHz) & 150 & 180 & 220 & 250 & 280 & 320 & 360\\\hline% \hline
%       Number of Bolometers & 2316 & 2316 & 3360 & 3202 & 3360  & 2648 & 3360\\\hline
%       NEQ/U ($\mu K \sqrt{s}$) & 4.17 & 4.36 & 5.35 & 6.19 & 8.81  & 13.69 & 23.33\\\hline
%       FWHM (') & 7.2 & 6.0 & 4.9 & 4.4 & 3.9  & 3.6 & 3.2\\\hline
%     \end{tabular}
%   \end{center}
%   \caption{The number of bolometers, sensitivity and resolution per frequency band for \ac{EBEX-IDS}.}
%   \label{tab:ebexidsinfo}
% \end{table}

We achieve polarization sensitivity by re-using the \ebex~sapphire half-wave plate in combination with polarization sensitive \ac{SAMP}, a technology already deployed by the ground-based telescopes \pb2~and \sptg.
\ac{EBEX-IDS} will operate 20,562 bolometers divided into 3,427 tri-chroic \ac{SAMP}s.
To read out these detectors, we will use \ac{FDM} readout with a multiplexing factor of 105 which will disspate 1.3~kW of power.
Increasing the multiplexing factor is essential for limiting the power dissipation.
We will cool the electronic boards with liquid coolant reaching the \ac{FPGA} in thermal contact with 6.7~m$^2$ of radiator panels.

In this paper, we discuss the challenges and progress to develop these pixels for balloon-borne platforms.
We develop low thermal conductance bolometers for optimal operation where 0.6~pW of power is expected to be absorbed by the bolometers at 150~GHz.
We also increase the multiplexing factor of the \ac{FDM} technology to 105, the highest multiplexing factor for the \ac{FDM} to date, to achieve the desired sensitivity.

\section{Low Thermal Conductance Detector Development}

\begin{figure}[htbp]
\begin{center}
\includegraphics[width=0.85\linewidth, keepaspectratio]{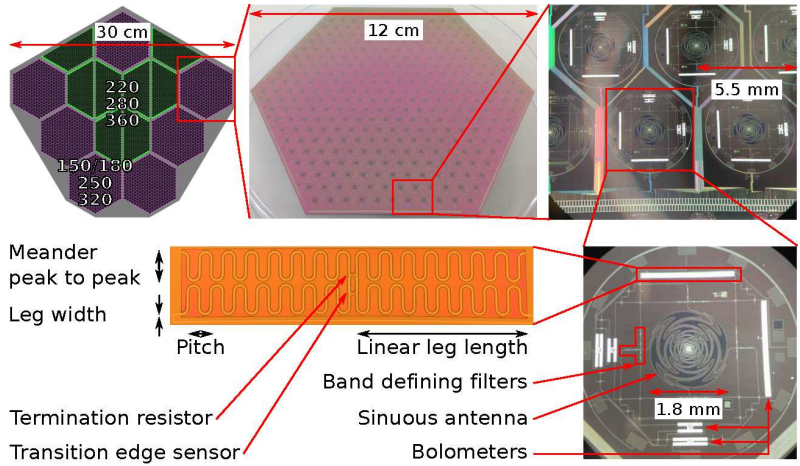}
\caption{
{\it Top left}: a model of the \ac{EBEX-IDS} focal plane composed of low frequency (150 or 180, 250, and~320 GHz) and high frequency pixels (220, 280, and~360 GHz).
{\it Top center}: a picture of an \ac{EBEX-IDS} bolometer wafer.
{\it Top right}: a picture of four \ac{SAMP}s.
{\it Bottom right}: a picture of a \ac{SAMP} tri-plexer pixel with straight bolometer legs.
{\it Bottom left}: a picture of a bolometer with four distinct meandered legs.
}
\end{center}
\label{fig:zigzagPanel}
\end{figure}

The \ac{SAMP} shown in Figure~\ref{fig:zigzagPanel} is fabricated with a series of sputter, \ac{PECVD}, and evaporation thin film deposition steps, each patterned with stepper and/or contact lithography.
Each metallic broadband sinuous antenna absorbs the incident light focused by an individual alumina lenslet.
We are developing a new anti-reflection coating method with thermal spray to maximize transmission~[2].
The antenna is composed of two perpendicular pairs of arms which are each sensitive to one polarization of light.
Niobium microstrips over silicon nitride couple the antennas to band-defining Dolph-Chebychev filters and to manganese doped aluminium \ac{TES} via a titanium termination resistor in strong thermal contact with the \ac{TES}.
The \ac{TES}s are coupled to the bath through weak thermal links.
Niobium leads, which provide the voltage bias to the \ac{TES}, connect to bond pads which are wire-bonded to custom \ac{LC} boards and the rest of the \ac{FDM} readout scheme.
Each bolometer encodes the signal from one frequency band and one polarization and each \ac{SAMP} encodes the signal from three bands and both polarizations.

We calculate the emission from the \ac{CMB}, the emission from the atmosphere, and the transmission and emission of each of the optical elements from first principles and derive the expected power absorbed by the bolometers~[3].
We calculate a value of 0.2~pW within the 150~GHz band, the frequency band for which we expect the smallest power, and include a 0.4~pW margin to conservatively account for the known challenge of assembling a cryostat to the theoretical specifications.
Assuming a safety factor of 2.5, we require the \ac{EBEX-IDS} bolometers observing at 150~GHz to have an average thermal conductance $\bar{G}$ of 9~pW/K.
We specify the bolometer legs to be shorter than 1.5~mm due to constraints on the physical size of the focal plane and the number of pixels.
The \ac{TES} of the $\sim$70~pW/K bolometers designed for the ground-based telescopes \pb~and \sptg~are physically supported by four straight legs [2, 4].
In order to decrease $\bar{G}$, we thin and elongate these legs.
We also use a meandered leg design, inspired by \spider, which allows for the elongation of the effective length of the legs while preserving the size of the pixel; see Figure~\ref{fig:zigzagPanel}~[5].

\begin{figure}[htbp]
\begin{center}
\includegraphics[height=4.3 cm, keepaspectratio]{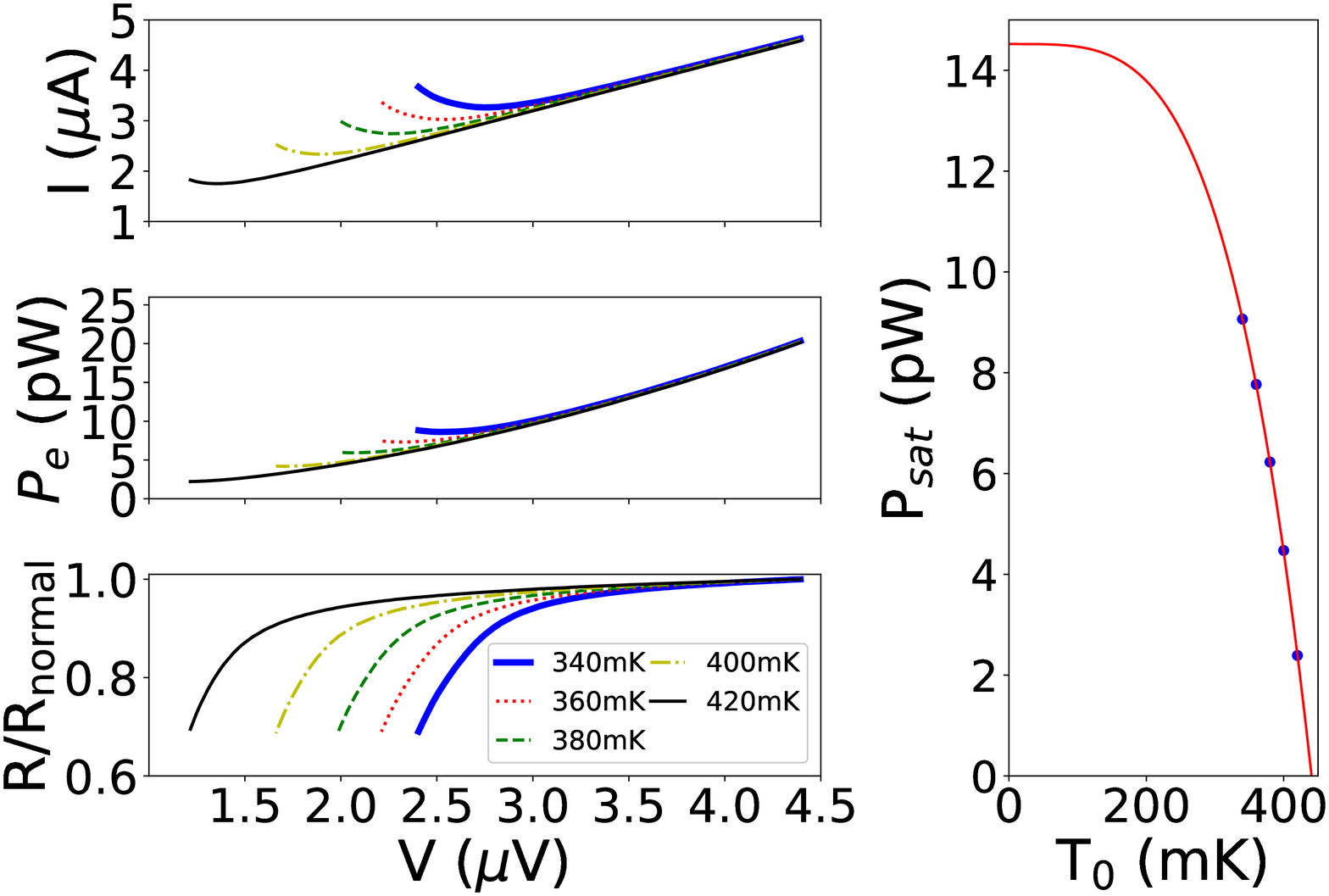}%\hspace{0.1\linewidth}
\includegraphics[height=4.3 cm, keepaspectratio]{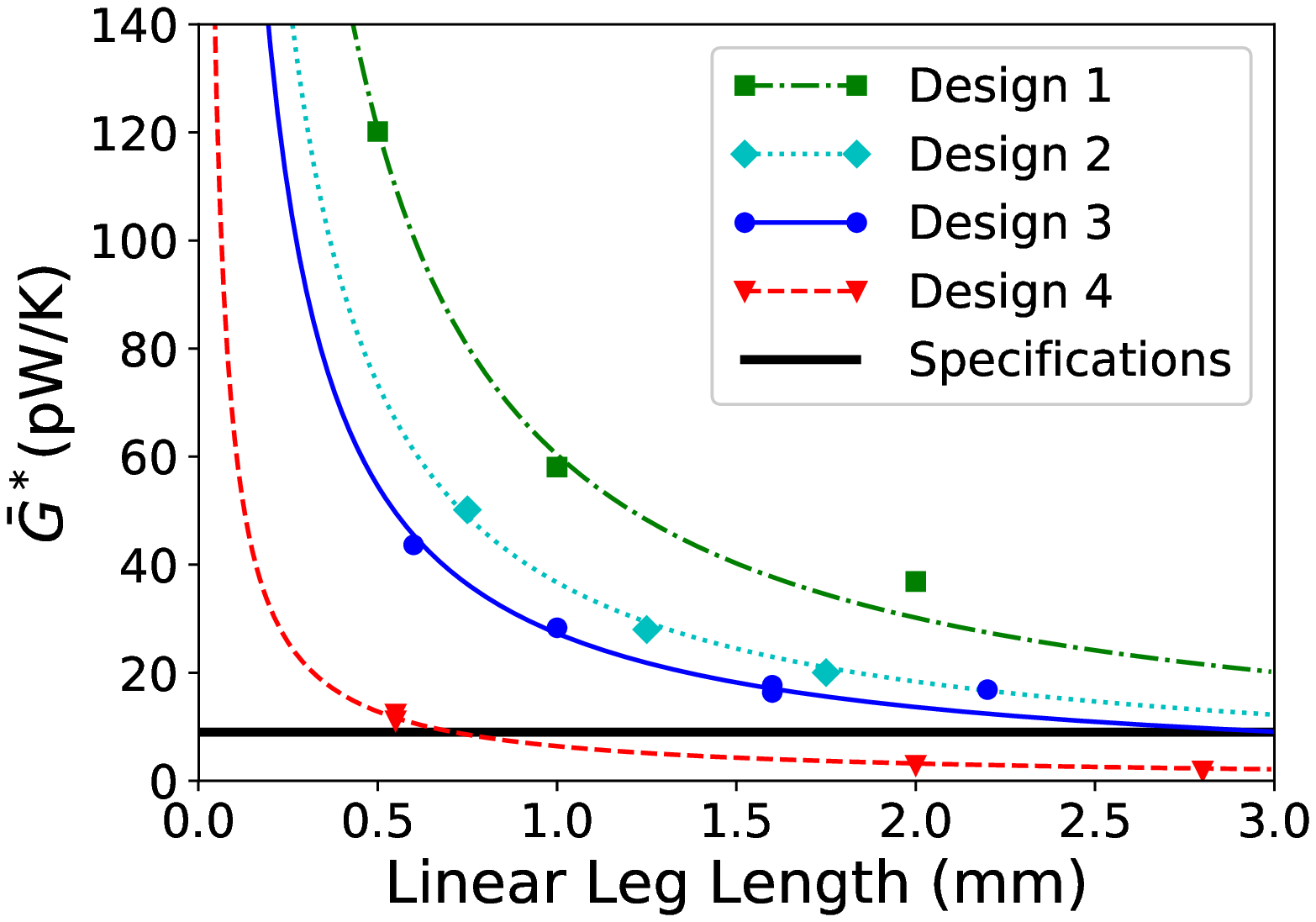}
\caption{
{\it Left}: the current $I$, the electrical power $P_e$ and the fraction of normal resistance $R/R_{normal}$ measured for a bolometer while it is dropped into its superconducting transition at various bath temperatures.
{\it Middle}: the saturation power as a function of the bath temperature ({\it blue circles}) fitted to the model described by Equation~\ref{eq:powerLaw} ({\it red solid}).
{\it Right}: the measured average thermal conductance extrapolated to \ac{EBEX-IDS} conditions ({\it symbols}) with the inverse leg length fitted model ({\it lines}) as a function of the linear leg length for the 
four designs described in Table~\ref{tab:designs}.
}
\label{fig:gbarVsLegLength}
\end{center}
\end{figure}

We measure the maximal power the bolometer can absorb before saturation $P_{sat}$, which depends on the temperature of the bath $T_0$ and the critical temperature of the \ac{TES} $T_c$, by measuring the electric power required to operate the detector in the strong electrothermal feedback regime in a dark environment and calculate 
\begin{equation}
\bar{G} (T_0,\ T_c) = P_{sat} (T_0,\ T_c) /  (T_c - T_0).
\label{eq:Gbar}
\end{equation}
A measurement of $P_{sat}$ at various bath temperatures is shown in Figure~\ref{fig:gbarVsLegLength}.
We also measure $T_c$ by providing a small voltage bias and measuring the resistance of the \ac{TES} as a function of the bath temperature~[3].
We assume the thermal conductivity of the bolometer follows a power law $\kappa = \kappa_0 T^n$ and hence
\begin{equation}
P_{sat} (T_0,\ T_c) = \frac{P_{sat,\ 0}}{n+1} \left( T_c^{n+1} - T_0^{n+1} \right),
\label{eq:powerLaw}
\end{equation}
where $P_{sat,\ 0}$ is the theoretical saturation power at 0~K.
We report the average thermal conductance in \ac{EBEX-IDS} conditions, which we denote with a $^*$.
We set $T_0^*$ to 250~mK by design and we specify $T_c^*$ to 440~mK to minimize the contribution of the phonon noise~[2].
We extract $n$ by measuring $P_{sat}$ at various bath temperature and we fit for $P_{sat,\ 0}$, $n$ and $T_c$.
We find an average value of 2.6 for $n$ and critical temperatures consistent with the measurement described previously.

We report on four leg designs for which the characteristic dimensions are defined in Figure~\ref{fig:zigzagPanel} and are quantified in Table~\ref{tab:designs}.
Two sets of chips with different characteristics were tested for design 4.
Initial designs with only two straight legs instead of four showed mechanical instability and were not further pursued. 
We measure that $\bar{G}$ is inversely proportional to the leg length and we tune $\bar{G}^*$ using this parameter; see Figure~\ref{fig:gbarVsLegLength}.
Design 4 with a projected critical temperature of 440~mK meets the 9~pW/K \ac{EBEX-IDS} specifications for bolometer legs longer than 0.71~mm .
The critical temperature will be decreased by reducing the fraction of aluminium in the \ac{TES} during fabrication.
The next design will focus on improving detector yield by implementing new release techniques to minimize the stress on the legs and by implementing small design changes to minimize asymmetries.

\begin{table}[htbp]
  \begin{center}
    \begin{tabular}{|c|c|c|c|c|c|c|c|} \hline
      Design & Leg type & Leg width & Meander peak to peak & Pitch & P$_{\mathrm{sat}}^{\dagger \ddagger}$ & T$_{\mathrm{c}}$ & $\bar{G}^{* \ddagger}$\\
             & & $\mu$m & $\mu$m & $\mu$m & pW & mK & pW/K\\\hline
      1 & Straight & 30, 18 & N/A & N/A & 13 & 500 & 40 \\\hline
      2 & Straight & 16 & N/A & N/A & 6.9 & 490 & 25 \\\hline
      3 & Meander & 16 & 200 & 150 & 6.5 & 510 & 18 \\\hline
      4.1 & Meander & 12 & 140 & 60 & 1.3 & 510 & 4.1 \\\hline
      4.2 & Meander & 12 & 140 & 60 & 0.6 & 400 & 4.4 \\\hline
      \multicolumn{8}{l}{\footnotesize$^{\dagger}$ corrected to $T_0^*$}\\
      \multicolumn{8}{l}{\footnotesize$^{\ddagger}$ for 1.5~mm legs}
    \end{tabular}
  \end{center}
  \caption{Physical characteristics of the bolometer designs.
    The dimensions are defined in Figure~\ref{fig:zigzagPanel}.
    Design 1 has two 30~$\mu$m wide legs supporting the microstrip and two 18~$\mu$m wide legs supporting the bias lines.
    Design 4 has its ground planes removed on the two legs carrying the bias lines.
}
  \label{tab:designs}
\end{table}

\begin{figure}[htbp]
\begin{center}
\includegraphics[height=3.7cm, keepaspectratio]{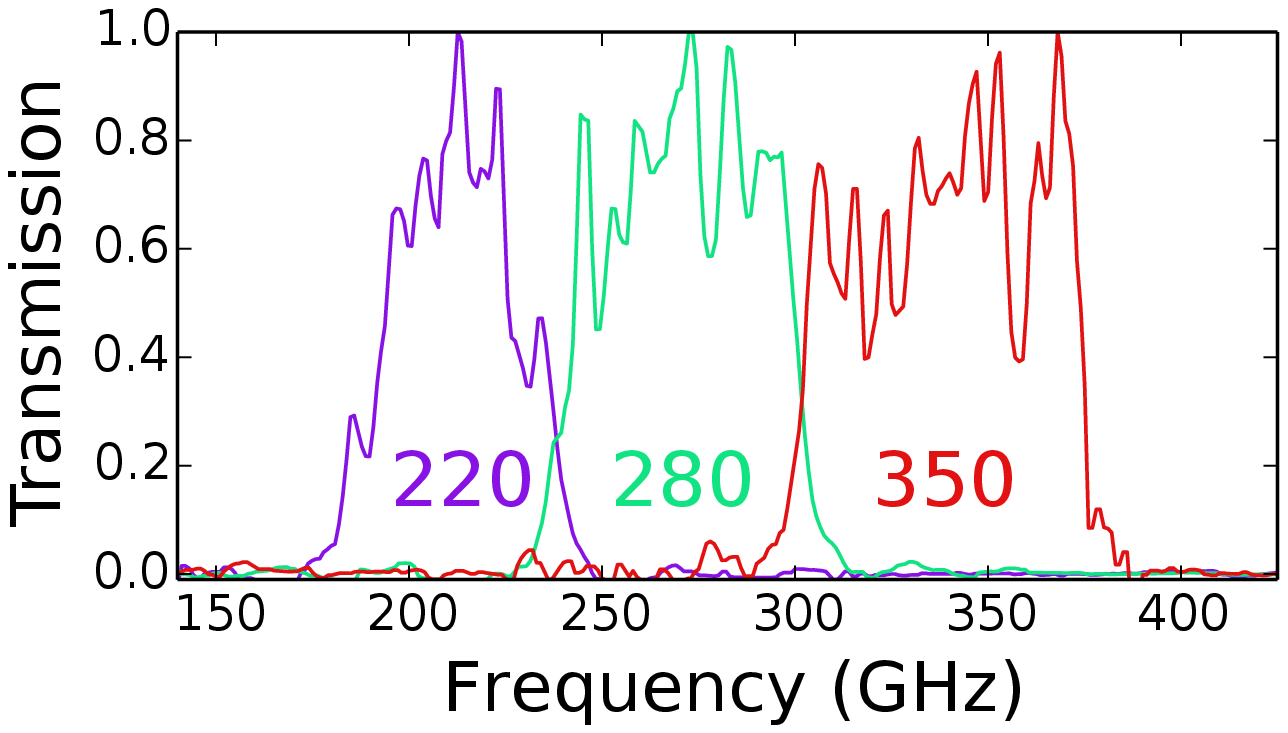}
\includegraphics[height=3.7cm, keepaspectratio]{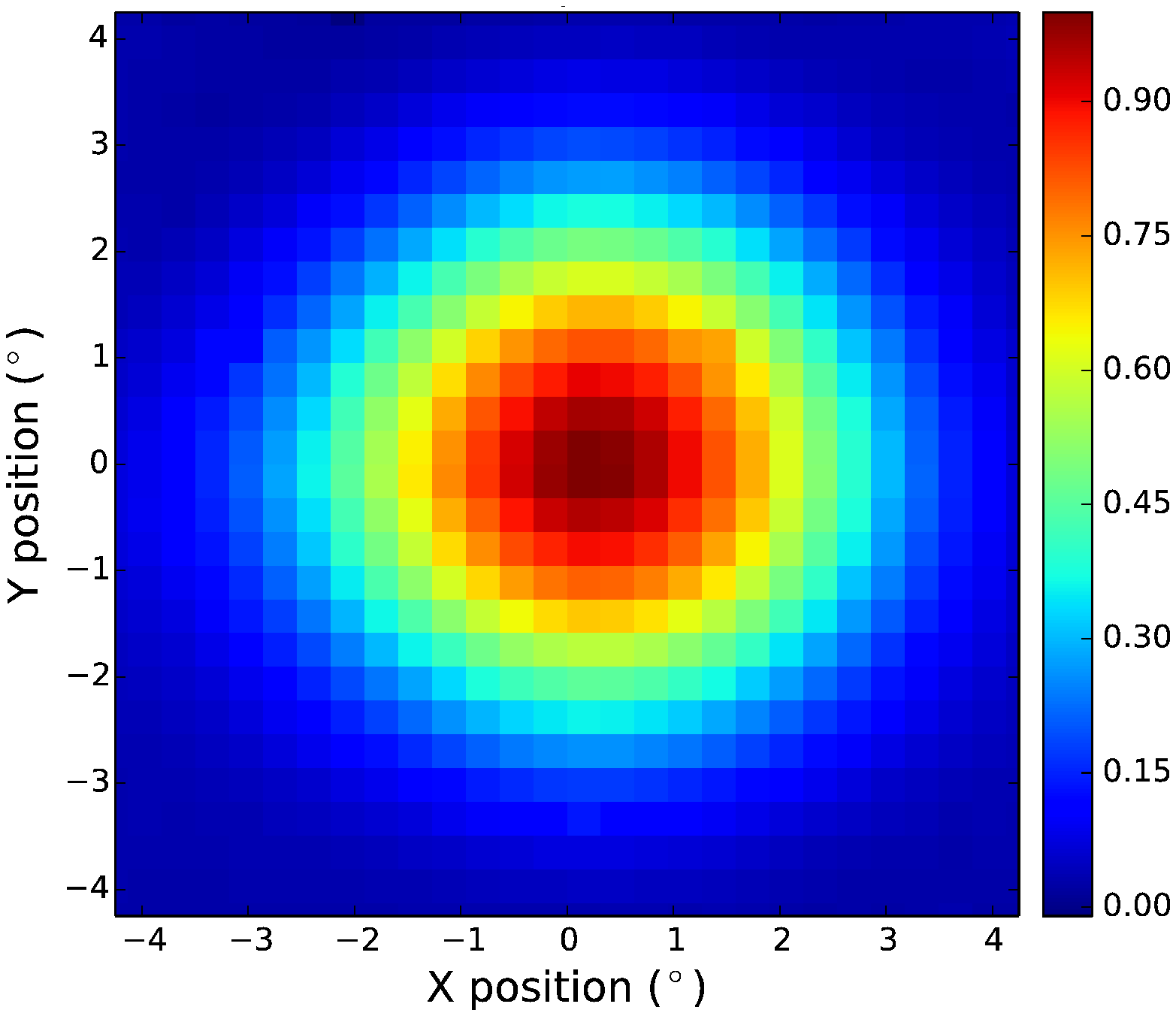}
\caption{
{\it Left}: the peak-normalized transmission of a 220, 280, and 350~GHz tri-plexer pixel close to the \ac{EBEX-IDS} specifications.
{\it Right}: the beam of a 220~GHz \ac{SAMP} pixel designed for \pb2~sharing the \ac{EBEX-IDS} optical design~[2, 6].
}
\label{fig:optics}
\end{center}
\end{figure}

Finally, the optical properties of a prototype with frequency bands close to \ac{EBEX-IDS} meet specifications as shown in Figure~\ref{fig:optics}.
We measure the transmission bands of a triplexer prototype pixel with a Fourier transform spectrometer using a ceramic heater as a thermal source.
The band edges are as expected from simulations~[6].
The beam is measured by translating a chopped liquid nitrogen source in an X-Y raster scan below a test cryostat and recording the response at each location.

\section{Readout}

\ac{EBEX-IDS} will use the latest \ac{FDM} technology: the ICE boards~[7].
With the \ac{FDM}, we simultaneously read out a group of detectors with only 2 wires.
The ICE boards support a multiplexing factor of $\times$128 and a multiplexing factor of $\times$68 has been demonstrated by \sptg~using the 1.6~-~5.3~MHz frequency range.
For \ac{EBEX-IDS} to achieve its planned sensitivity, a multiplexing factor of $\times$105 is required.
We increase the multiplexing factor through two steps.

\begin{figure}[htbp]
\begin{center}
\includegraphics[width=0.95\linewidth, keepaspectratio]{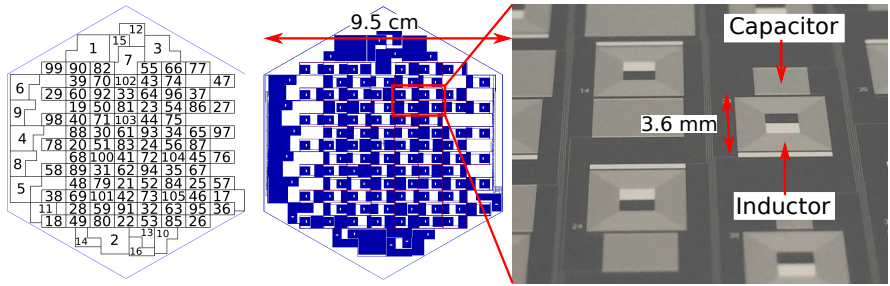}
\caption{
{\it Left}: the location of the \ac{LC} pairs on the \ac{LC} chip. The lowest frequency is labelled ``1'' and the highest, ``105''.
{\it Middle}: a model of the \ac{LC} chip.
{\it Right}: a zoom on a picture of \ac{LC} pairs.
}
\label{fig:LCchip}
\end{center}
\end{figure}

First, we decrease the spacing of the resonant frequencies by increasing the inductance of the band selecting inductors from 60 to 90~$\mu$H.
The crosstalk due to the mutual inductance of the inductors can be arbitrarily decreased by ensuring the \ac{LC}s of neighbouring frequencies are physically far on the \ac{LC} chip; see Figure~\ref{fig:LCchip}.
The magnitude of crosstalk due to voltage biases from the frequency neighbours is given by $\left| \frac{R}{2 \Delta \omega L} \right|^2$.
We calculate crosstalk of 0.09\% (0.05\%) at 0.4~MHz (5.3~MHz) with a 30~kHz (112~kHz) spacing.
The magnitude of the crosstalk caused by the signal from a bolometer heating its neighbour due to voltage drop across the non-zero impedance of the wiring or the \ac{SQUID} is $X_I \frac{L_{stray}}{L} \frac{\omega}{\Delta \omega}$, where $X_I$ is the ratio between the current through bolometer $i$~$\pm$~1 caused by bolometer $i$ to the current through bolometer $i$.
We calculate crosstalk of 0.05\% (0.04\%) at 0.4~MHz (5.3~MHz) with a 30~kHz (112~kHz) spacing and a stray inductance $L_{stray}$ of 100~$\mu$H~[8].
The electrical crosstalk remains well below the expected optical crosstalk of 1\%.

Second, we extend the range of frequency used.
Using frequencies above 5.3~MHz is challenging because of capacitive strays from the \ac{FDM} band selecting capacitors and inductive strays from wiring; both increase with frequency.
We therefore extend the frequency range towards lower frequencies to 0.4~-~5.3~MHz.

We are now developing a new \ac{LC} chip fabricated with a 2~$\mu$m lithographic process.
The inductors are 3.6~$\times$~3.6~mm squares composed of 200 turns of 3~$\mu$m wide niobium wires also separated by 3~$\mu$m.
We use 10~-~1759~pF interdigitated capacitors to define the bolometer bias frequency. 
Figure~\ref{fig:LCchip} shows the design of the \ac{EBEX-IDS} chip where the 105 \ac{LC} pairs are position on a hexagonal silicon wafer with 5.5~cm sides, similar in size to the hexagonal bolometer wafer with 5~cm sides.
We use a frequency spacing of 30~kHz for the frequency range 0.4~-~1.5~MHz and a logarithmic spacing for the frequency range 1.6~-~5.3~MHz, which is 30~kHz at 1.5~MHz and 112~kHz at 5.3~MHz.

A hexagonal \ac{PCB} mounting board for the \ac{LC} chip with 6~cm sides is being designed.
This board will be stacked under the bolometer wafer for compactness, hence the hexagonal shape.

\section{Summary}

We have fabricated and characterized bolometers that exceed the \ac{EBEX-IDS} specifications.
Bolometers composed of four 12~$\mu$m wide, 140~$\mu$m peak to peak, and 0.71~mm long meandered legs with a 60~$\mu$m pitch, as defined in Figure~\ref{fig:zigzagPanel}, in combination with \ac{TES}s which have a 440~mK critical temperature meet the 9~pW/K specified average thermal conductance. 
We are in the process of fabricating optical pixels for \ac{EBEX-IDS}.
\ac{EBEX-IDS} will use the ICE boards to read out detectors in the frequency domain with a multiplexing factor 60\% higher than the highest factor to date.
The design of the sub-Kelvin readout components is almost completed and will soon be tested.
The lower thermal conductance bolometers achieved with the meandered leg design combined with the highest \ac{FDM} factor to date increases the readiness of this detectors technology for balloon platform and will allow \ac{EBEX-IDS} to achieve its specified sensitivity.

\begin{acknowledgements}
The EBEX-IDS Collaboration would like to thank the support from NASA (NNX17AH30G).
\end{acknowledgements}

\pagebreak

\begin{acronym}
    %A
    \acro{ACS}{attitude control system}
    \acro{ADC}{analog-to-digital converters}
    \acro{ADS}{attitude determination software}
    \acro{AHWP}{achromatic half-wave plate}
    \acro{AMC}{Advanced Motion Controls}
    \acro{ARC}{anti-reflection coating}
    \acro{ATA}{advanced technology attachment}
    %B
    \acro{BRC}{bolometer readout crates}
    \acro{BLAST}{Balloon-borne Large-Aperture Submillimeter Telescope}
    %C
    \acro{CANbus}{controller area network bus}
    \acro{CMB}{cosmic microwave background}
    \acro{CMM}{coordinate measurement machine}
    \acro{CSBF}{Columbia Scientific Balloon Facility}
    \acro{CCD}{charge coupled device}
    %D
    \acro{DAC}{digital-to-analog converters}
    \acro{DASI}{Degree~Angular~Scale~Interferometer}
    \acro{dGPS}{differential global positioning system}
    \acro{DfMUX}{digital~frequency~domain~multiplexer}
    \acro{DLFOV}{diffraction limited field of view}
    \acro{DSP}{digital signal processing}
    %E
    \acro{EBEX}{E~and~B~Experiment}
    \acro{EBEX2013}{EBEX2013}
    \acro{EBEX-IDS}{E~and~B~Experiment - Inflation and Dust Surveyor}
    \acro{ELIS}{EBEX low inductance striplines}
    \acro{EP1}{EBEX Paper 1}
    \acro{EP2}{EBEX Paper 2}
    \acro{EP3}{EBEX Paper 3}
    \acro{ETC}{EBEX test cryostat}
    %F
    \acro{FDM}{frequency domain multiplexing}
    \acro{FPGA}{field programmable gate arrays}
    \acro{FCP}{flight control program}
    \acro{FOV}{field of view}
    \acro{FWHM}{full width half maximum}
    %G
    \acro{GPS}{global positioning system}
    %H
    \acro{HPE}{high-pass edge}
    \acro{HWP}{half-wave plate}
    %I
    \acro{IA}{integrated attitude}
    \acro{IP}{instrumental polarization} 
    %J
    \acro{JSON}{JavaScript Object Notation}
    %L
    \acro{LDB}{long duration balloon}
    \acro{LED}{light emitting diode}
    \acro{LCS}{liquid cooling system}
    \acro{LC}{inductor and capacitor}
    \acro{LPE}{low-pass edge}
    %M
    \acro{MLR}{multilayer reflective}
    \acro{MAXIMA}{Millimeter~Anisotropy~eXperiment~IMaging~Array}
    %N
    \acro{NASA}{National Aeronautics and Space Administration}
    \acro{NDF}{neutral density filter}
    %P
    \acro{PCB}{printed circuit board}
    \acro{PE}{polyethylene}
    \acro{PECVD}{plasma-enhanced chemical vapor deposition}
    \acro{PTFE}{polytetrafluoroethylene}
    \acro{PME}{polarization modulation efficiency}
    \acro{PSF}{point spread function}
    \acro{PV}{pressure vessel}
    \acro{PWM}{pulse width modulation}
    %R
    \acro{RMS}{root mean square}
    %S
    \acro{SAMP}{sinuous antenna multichroic pixels}
    \acro{SLR}{single layer reflective}
    \acro{SMB}{superconducting magnetic bearing}
    \acro{SQUID}{superconducting quantum interference device}
    \acro{SQL}{structured query language}
    \acro{STARS}{star tracking attitude reconstruction software}
    %T
    \acro{TES}{transition edge sensors}
    \acro{TDRSS}{tracking and data relay satellites}
   \acro{TM}{transformation matrix}

\end{acronym}

\end{document}